\documentstyle[alife8,amssymb,epsfig,pstricks]{article}
\title{Diversity Evolution}
\author{\aindx{Russell K. Standish}{Standish}\\
School of Mathematics, University of New South Wales\\
2052, Sydney, Australia\\
R.Standish@unsw,.edu.au\\
http://parallel.hpc.unsw.edu.au/rks
}

\newcommand{\br}{\mbox{\boldmath{$r$}}}          
\newcommand{\bbeta}{\mbox{\boldmath{$\beta$}}}   
\newcommand{\bgamma}{\mbox{\boldmath{$\gamma$}}} 
\newcommand{\bmu}{\mbox{\boldmath{$\mu$}}}       
\newcommand{\bn}{\mbox{\boldmath{$n$}}}          
\newcommand{\Acum}{\bar{A}_{\mbox{\rm\scriptsize cum}}}
\newcommand{\Anew}{A_{\mbox{\rm\scriptsize new}}}
\newcommand{\EcoLab}{{\sffamily\slshape
    \mbox{\raisebox{.5ex}{Eco}\hspace{-.4em}{\makebox[.5em]{L}ab}}}}

\newlength{\figwidth}
\begin{document}
\maketitle

\begin{abstract}
  Bedau\index{Bedau-Packard evolutionary statistics}
  \index{diversity}\index{evolution!of diversity} has developed a
  general set of evolutionary statistics that quantify the adaptive
  component of evolutionary processes. On the basis of these measures,
  he has proposed a set of 4 classes of evolutionary system. All
  artificial life sytems so far looked at fall into the first 3
  classes, whereas the biosphere, and possibly the human economy
  belongs to the 4th class. The challenge to the artificial life
  community is to identify exactly what is difference between these
  natural evolutionary systems, and existing artificial life systems.
  
  At ALife VII, I presented a study using an artificial evolutionary
  ecology called \EcoLab.\index{EcoLab} Bedau's statistics captured
  the qualitative behaviour of the model. \EcoLab{} exhibited
  behaviour from the first 3 classes, but not class 4, which is
  characterised by unbounded growth in diversity. \EcoLab{} exhibits a
  critical surface given by an inverse relationship between
  connectivity and diversity, above which the model cannot tarry long.
  Thus in order to get unbounded diversity increase, there needs to be
  a corresponding connectivity reducing (or food web\index{food web}
  pruning) process.  This paper reexamines this question in light of
  two possible processes that reduce ecosystem\index{ecosystem}
  connectivity: a tendency for specialisation\index{specialization}
  and increase in biogeographic\index{biogeography} zones through
  continental drift.
\end{abstract}

\section{Introduction}

During the Phanerozoic\index{Phanerozoic} (540Mya--present), the
diversity of the biosphere\index{biosphere} (total number of species,
also known as biodiversity) has increased dramatically. The trend is
most clear for intermediate taxonomic levels (families and orders), as
fossil species data is too incomplete and higher taxonomic levels
(phylum and class diversity) have been fairly constant since the
Paleozoic. A recent review is given by
Benton\citeyear(2001){Benton01}. The most completely documented
diversity trend is amongst marine animals, which exhibits a plateau
during the Paleozoic\index{Paleozoic plateau} (540-300Mya), followed
by an accelerating diversity curve since the end of the Permian. The
corresponding trend amongst continental, or land animals is
characterised by a clear exponential growth since the first species
colonised dry land during the Ordovician. Benton argues that the
terrestrial trend is more characteristic than the marine trend, owing
to the far greater diversity shown amongst land animals, even though
the marine fossil record is more complete. Similar trends have been
reported for plants\cite{Tiffney-Niklas90}.

Bedau {\em et al.}\citeyear(1998){Bedau-etal98} introduced a couple of
measures to capture the amount of adaptation happening in a general
evolutionary system. The basic idea is to compare the dynamics of the
system with a neutral shadow system in which adaptation is destroyed by
randomly mixing adaptive benefits amongst the components of the system
(think of the effects ultra-Marxism might have on an economy!). The
amount of adaptive activity (numbers of each component in excess of
the shadow model integrated over time) and adaptive creativity
(numbers of speciations per unit time exceeding a threshold of
activity) is measured. Bedau has also introduced a general neutral
shadow model that obviates the need to generate one on a case by case
basis\cite{Rechtsteiner-Bedau99}.

Using these measures, it is possible to distinguish 3 classes of
activity:
\begin{enumerate}
\item unadaptive evolution, when the mutation rate is so high that
  organisms have insufficient time to have their adaption tested
  before being killed off by another mutation
\item adapted but uncreative evolution, when species are highly
  adapted, but mutation is so low that ecosystems remain in perpetual
  equilibrium
\item creative, adaptive evolution, when new species continuously enter
  the system, and undergo natural selection
\end{enumerate}

The Biosphere appears to be generating open ended novelty --- not only
is it creative, but it is {\em unboundedly} creative. Evidence for
this exists in the form of the intricate variety of mechanisms with
which different organisms interact with each other and the
environment, and also in the sheer diversity of species on the planet.
Whilst there is no clear trend to increasing organismal
complexity\index{complexity}\cite{McShea96}, there is the clear trend
to increasing diversity mentioned above, which is likely to be
correlated with ecosystem complexity. Bedau takes diversity as a third
evolutionary measure, and distinguishes between {\em bounded} and {\em
  unbounded creative} evolution, according to whether diversity is
bounded or not. All artificial evolutionary systems examined to date
have, when creative, exhibited bounded
behaviour\footnote{Channon\citeyear(2001){Channon01} claims his
  Geb\index{Geb} artificial life system exhibits unbounded creative
  behaviour}. This was also the case of the \EcoLab{} model
\cite{Standish00c}. Bedau has laid down a challenge to the artificial
life community to create an unbounded, creative evolutionary system.

\section{Ecosystem Complexity}

The heart of the idea of unbounded creative evolutionary activity is
the creation and storage of information. The natural measure of this
process is {\em information based complexity}, which is defined in the
most general form in\cite{Standish01a}. The notion, drawing upon
Shannon entropy\index{entropy!Shannon} and Kolmogorov
complexity\index{complexity!Kolmogorov}\cite{Li-Vitanyi97} is as
follows:

A language ${\cal L}_1=(S,\mu)$, is a countable set of possible
descriptions $S$, and a map $\mu:S\rightarrow\{0,1\}$. We say that
$s,s'\in S$ have the same meaning iff $\mu(s,s')=1$. Denote the length
of $s$ as $\ell(s)$ and $S_n=\{s\in S:\ell(s)=n\}$. The information
content (or {\em complexity}) of a description $s$ is given by:
\begin{equation}\label{complexity}
{\cal C}(s) = -\lim_{n\rightarrow\infty} \log_2 \frac{
\mathrm{card}(\{s'\in S_n:\mu(s,s')=1\})}{\mathrm{card}(S_n)} 
\end{equation}
In the usual case where the interpreter (which defines $\mu$) only
examines a finite number of symbols to determine a string's meaning,
$C(s)$ is bounded above by $\ell(s)\log_2 B$ where $B$ is the size of
the alphabet. This is equivalent to the notion of {\em prefix codes}
in algorithmic information theory.

Now consider how one might measure the complexity of an ecosystem.
Diversity is like a count of the number of parts of a system --- it is
similar to measuring the complexity of a motor car by counting the
number of parts that make it up. But then a junkyard of car parts has
the same complexity as the car that might be built from the parts. In
the case of ecosystems, we expect the interactions between species to
be essential information that should be recorded in the complexity
measure. But a simple naive counting of food web connections is also
problematic, since how do we know which connections are significant to
a functioning ecology?

To put the matter on a more systematic footing, consider a tolerance
$\varepsilon$ such that $x, y \in \reals$ are considered identical if
$|x-y|<\varepsilon$. Now two different population dynamics $\dot
x=f(x)$ and $\dot x=f'(x)$, where 
\begin{displaymath}
x\in\reals^{n+}\equiv\{x\in\reals^n:x_i\geq0\},
\end{displaymath}
 can be considered identical
(i.e. $\mu(f,f')=1$ iff\footnote{As an anonymous referee pointed out,
  trajectories decsribed by $f$ and $f'$ may diverge exponentially in
  time, and that a better definition of equivalence would also require
  similarity of the attractor sets as well. The results derived here
  would only be a lower bound of the ecosystem complexity under this
  more refined definition of equivalence.}
\begin{equation}\label{defeq}
|f(x)-f'(x)|_\infty<\varepsilon,\, \forall
x\in\reals^{n+}.
\end{equation}

At this point for the sake of concreteness, let us consider
Lotka-Volterra\index{Lotka-Volterra} dynamics:
\begin{equation}
\dot {\bf x} = {\bf r}*{\bf x}+{\bf x}*\beta {\bf x}
\end{equation}
where $*$ refers to elementwise multiplication, ${\bf r}$ is the net
population growth rate and $\beta$ is the matrix of interspecific
interaction terms.

Over evolutionary time, the growth coefficients $r_i$, the
self-interaction coefficients $\beta_{ii}$ and the interspecific
interaction coefficients $\beta_{ij}, i\neq j$ form particular
statistical distributions $p_r(r_i), p_d(\beta_{ii})$ and
$p_o(\beta_{ij})$ repectively.

Since inequality (\ref{defeq}) must hold over all of the positive cone
$\reals^{n+}$, it
must hold for population density vectors $|x|\ll 1$ and $|x|\gg
1$. In which case eq. (\ref{defeq}) can be broken into independent
component conditions on ${\bf r}$ and $\beta$ can be written:
\begin{eqnarray}
|r_i-r'_i|&\leq& \varepsilon, \, \forall i\\
||\bbeta-\bbeta'||_\infty&\leq&\varepsilon.
\end{eqnarray}

Since these conditions are independent, they contribute additively to
the overall complexity (\ref{complexity}). The term for the growth
coefficients is given by:
\begin{eqnarray}
{\cal C}_r &=& -\log_2 \prod_i
\int_{|r_i-r'_i|\leq\varepsilon}p_r(r'_i)dr'_i\nonumber\\
&\approx&  -\sum \log_2 p_r(r_i) - D\log_2 2\varepsilon
\end{eqnarray}
where $\varepsilon\ll 1$, and $D$ is the ecosystem diversity.

The complexity term for the interaction terms is given by
\begin{eqnarray}
{\cal C}_\beta &=&
-\log_2\int_{\sum_j|\beta_{ij}-\beta'_{ij}|<\varepsilon}
\prod_{i\neq
j}p_o(\beta'_{ij})\prod_ip_d(\beta'_{ij})\prod_{i,j}d\beta'_{ij}
 \nonumber\\
&\approx& \sum_{i\neq j}\log_2p_o(\beta_{ij}) +
\sum_i\log_2p_d(\beta_{ii}) + \nonumber\\
&& D^2\log_2 2\varepsilon - 1\label{C_beta}
\end{eqnarray}

If $\varepsilon$ is chosen very small, the total ecosystem complexity
is proportional to $D^2$. This is because the zeros of the interaction
matrix are encoding information. However, if
$\varepsilon=\frac1{2p_o(0)}$, then (\ref{C_beta}) becomes:
\begin{equation}
{\cal C}_\beta=D^2C\langle\log_2p_o(\beta_{ij})\rangle +
D\langle\log_2p_d(\beta_{ii})\rangle+o(D)
\end{equation}
This gives flesh to our intuitive notion that complexity should
somehow be proportional to the number of connections making up the
food web.

Empirically, Lotka-Volterra dynamics has been shown to exhibit an
inverse relationship between connectivity and diversity $D\propto
C^{-1}$\cite{Standish98c}. May\citeyear(1972){May72} demonstrated this
relationship in connection with dynamical stability. However, it seems
unlikely that an ecology undergoing evolution is often stable.
If this result holds more generally, it implies that complexity is directly
proportional to diversity, so that diversity indeed is a good proxy
for ecosystem complexity. Although earlier foodweb studies
demonstrated this hyperbolic diversity-connectivity relationship, more
recently collected data suggests a relationship of the form $D\propto
C^{-1+\epsilon}$, with
$\epsilon\approx$0.3--0.4 \cite{Drossel-McKane02}. If complexity
indeeds scales superlinearly with diversity as suggested by latter
data, then a system displaying open-ended diversity growth is indeed
growing in complexity, however a system displaying bounded diversity
growth may still be growing in complexity.

\section{\protect\EcoLab}

\EcoLab{} is an evolutionary ecology, and is to my knowledge the first
published account of population dynamics being linked to an
evolutionary algorithm\cite{Standish94}. The next model to be
developed in this genre is Webworld\cite{Drossel-etal01}, which features a
more realistic ecological dynamics, and handles resource flow issues
better. Other models in this genre have appeared recently
\cite{Christensen-etal02,Anastasoff00}.

\EcoLab{} is also the name of a software package used for implementing
this model, as well as other models. The software is available from
\htmladdnormallink{http://parallel.hpc.unsw.edu.au/rks/ecolab}
{http://parallel.hpc.unsw.edu.au/rks/ecolab}. 

The model consists of Lotka-Volterra ecology:
\begin{equation}
\dot{\bn} = \br*\bn + \bn*\bbeta\bn + {\tt mutate}(\bmu,\br,\bn) + \bgamma*\nabla^2\bn.
\end{equation}
\bn\ is the population density vector, \br\ the growth rates (net
births-deaths in absence of competition), \bbeta\ the interaction
matrix, \bmu\ the (species specific) mutation rates and \bgamma\ the
migration rate. In the panmictic case, the \bgamma\ term is left out,
and \bn\ refers to total populations, rather than population
densities.

The mutation operator randomly adds new species $i$ into the system with
phenotypic parameters ($r_i$, $\beta_{ij}$, $\mu_i$ and $\gamma_i$)
varied randomly from their parent species. A precise documentation of
the mutation operator can be found in the \EcoLab{} Technical
Report\cite{Ecolab-Tech-Report}.

The \bn{} vector has integral valued components --- in assigning a
real valued vector to it, the values are rounded up {\em randomly}
with probability equal to the fractional part. For instance, the value
2.3 has a 30\% probability of being rounded up to 3, and a 70\%
probability of being rounded down. Negative values are converted to
zero. If a species population falls to zero, it is considered extinct,
and is removed from the system. It should be pointed out that this is
a distinctly different mechanism than the threshold method usually
employed to determine extinction, but is believed to be largely equivalent.

Diversity $D$ is then simply the number of species with $n_i>0$, and
connectivity is the proportion of interspecific connections out of all
possible connections:
\begin{displaymath}
C=\frac{1}{D^2}\sum_{i,j|\beta_{ij}\neq0,\, n_i>0,\, \& n_j>0}
\end{displaymath}

Spatial \EcoLab{} is implemented as a spatial grid, with the $\nabla^2$
term being replaced by the usual 5-point stencil.

\section{Specialisation}

A {\em specialist} is a species that only depends on a restricted
range of food sources, as opposed to a {\em generalist} which might
depend on many food sources. A specialist has fewer incoming
predator-prey links in the food web than does a generalist. Much
evolutionary variety is expressed in sophisticated defence mechanisms
that serve to suppress outgoing predator-prey links. In this context,
I will use the term {\em specialist} in a more general sense to refer
to species with a small number of food web links. In order for the
panmictic EcoLab model to generate an increasing diversity trend, a
corresponding specialisation trend must also be present (which it
isn't in the case of the usual mutation operator). Interestingly,
specialisation is usually considered to be the default mode of
evolution\cite{Vermeij87}. Generalists only exist because they happen to
be more robust against environmental perturbation.

This experiment involves modifying the mutation operator to bias it
towards removing interaction terms. The usual \EcoLab{} mutation operator
operator adds or removes connections according to $\lfloor1/r\rfloor$,
where $r\in(-1,1)$ is a uniform random
variate\citeyear(\EcoLab{} Technical Report){Ecolab-Tech-Report}. In
this experiment, a new 
experimental parameter $g\in(-1,1)$ (\verb+gen_bias+) is introduced such that
$r\in(-1+g,1+g)$ and the number of connections added or deleted is
given by $\lfloor (1+\mathrm{sgn}(r)g)/r\rfloor$. By specifying a very
negative value of $g$, the mutation operator will tend to produce
specialists more often than generalists. The code for this experiment
is released as \EcoLab{} 4.2.

\begin{figure}
  \epsfxsize=\figwidth\epsfbox{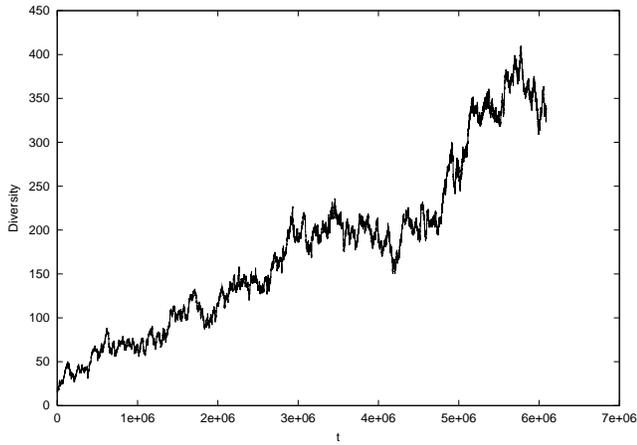}
\caption{Diversity growth for a typical run with $g=-0.9$}
\label{Diversity}
\end{figure}

\begin{figure}
  \epsfxsize=\figwidth\epsfbox{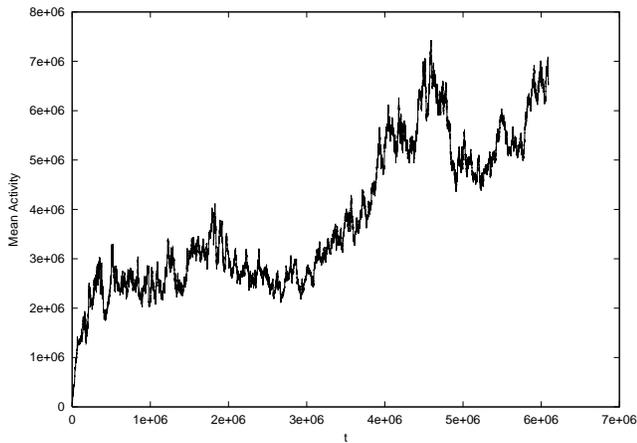}
\caption{Mean cumulative activity $\Acum(t)$}
\label{Activity}
\end{figure}

\begin{figure}
  \epsfxsize=\figwidth\epsfbox{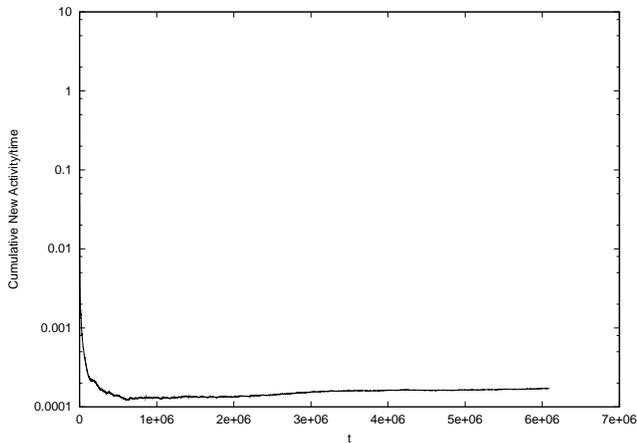}
\caption{Cumulative New Activity over time $\frac1t\int_0^t\Anew(t)dt$}
\label{New}
\end{figure}

\begin{figure}
  \epsfxsize=\figwidth\epsfbox{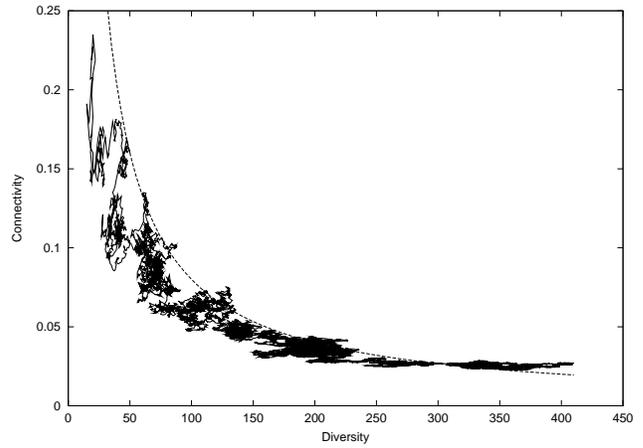}
\caption{Connectivity vs Diversity for the same run depicted in Figs. \ref{Diversity}--\ref{New}. The dashed curve is $8/D$.}
\label{Connectivity}
\end{figure}

\begin{figure}
\begin{pspicture}(-1,-.25)(5,3)
\psline{->}(0,0)(0,3)
\psline{->}(0,0)(5,0)
\psline(0,0)(3,3)
\psline[linestyle=dashed](2,0)(2,3)
\psline{<-}(2,1)(3,1)
\rput[l](3,1){Max $D$}
\pscurve(0,1.75)(2,2)(5,3)
\rput(-.5,1.5){$\frac1{C(D)}$}
\rput(2.5,-.5){$D$}
\end{pspicture}
\caption{Diversity is constrained to lie under the curve $1/C(D)$. The
  intersection of this curve with line $y=D$ gives the maximum
  possible diversity in the ecosystem. If $C(D)\leq o(D^{-1})$, then
  diversity is unbounded.}
\label{div-conn}
\end{figure}
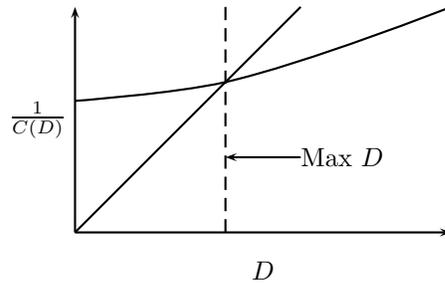

A typical run with $g=-0.9$ is shown in Figs.
\ref{Diversity}--\ref{Connectivity}. As described in
\cite{Standish00c}, activity is weighted by the population density,
not just presence of a particular species.  The results show unbounded
creative evolutionary activity (Class 4 behaviour). As can be seen
from Fig. \ref{Connectivity}, the system remains close to the
hyperbolic critical surface, yet the dynamic balance has been removed
by the specialisation trend. If we assume that $D\leq C_0/C(D)$, then
unbounded diversity growth can only happen if $C$ vanishes at least as
fast as $1/D$ (see Fig. \ref{div-conn}).  An ecosystem consisting
entirely of specialists has a constant number of foodweb links per
species, or $C\propto 1/D$. The presence of generalists in the
ecosystem damps the growth in diversity, and unbounded growth is only
possible if the proportion of generalists continually diminishes over
time.

\section{Continental Drift}

In \cite{Standish00c}, I suggested that one possible explanation for
the diversity growth since the end of the Permian was the breakup of
the supercontinent Pangaea.\index{Pangaea} A simple estimate given in
that paper indicated that the effect might account for a diversity
growth of about 3.5 times that existing during the Permian. This was
remarkably similar to the growth reported by\cite{Benton95}, however
it is worth noting that Benton's data referred to families, not
species. It is expected that the numbers of species {\em per family}
also increased during that time\cite{Benton01}. Furthermore, when
continental organism are included, familial diversity today is more
like 5 times the diversity during the Permian.

Unbeknownst to me at the time, Vallentine\citeyear(1973){Vallentine73}
had proposed essentially the same theory, called {\em biogeographic
  provincialism}\index{biogeographic provincialism} (the notion that
the number of biological provinces is increased through rearrangement
of the continents).\index{continental drift} The idea received some
serious support by Signor\citeyear(1990){Signor90}, although in a
later review he was less enthusiastic\cite{Signor94}.  Tiffney and
Niklas\citeyear(1990){Tiffney-Niklas90} examined plant diversity in
the northern hemisphere and concluded that plant diversity correlated
more with the land area of lowlands and uplands, rather than
continental breakup.  Benton\citeyear(1990){Benton90} is
characteristically sceptical of biogeographic provicialism as an
explanation of the diversity trend through the Phanerozoic.
Biogeography theory depends on an assumed dynamic balance between
speciation and extinction\footnote{Benton calls this a {\em dynamic
    equilibrium}, although it is nothing like what the term
  equilibrium denotes in dynamical system theory, and characterises
  biogeograhic theories as {\em equilibrial}}, which appears to be
contradicted by the fossil data for continental
animals\cite{Benton01}, which shows a strong exponential increase in
diversity through the Phanerozoic.

Since the \EcoLab{} model has this dynamic balance between speciation
and extinction when the dynamics self-organise to the critical surface
$D\approx(Cs^2)^{-1}$, I experimented with the spatial version of
\EcoLab{} reported in \cite{Standish98c}. The maximum migration rate
$|\bgamma|_\infty$ was
swept up and down exponentially in time according to $0.9^{t/1000}$,
{\em i.e.} with a time constant of about 9500 timesteps, by scaling
\bgamma{} by 0.9 every 1000 timesteps (and then inverting the scaling
factor every 174,000 timesteps). It is a little hard
to relate \EcoLab{} figures to biological evolution. The maximum growth
rate in \EcoLab{} is $0.01$, so the doubling time for the fastest
organism in the ecosystem is around 100 timesteps. This might
correspond to a year or so of real time. So migration rates are being
forced much faster than is typical in the real world. However, in
EcoLab we also tend run the mutation rate quite high, with adaptive
speciations happening every 1000 timesteps or so. If the mutation rate
is too high, natural selection has no chance to weed out non-adaptive
species, if too low, too much computing resource is need to obtain
interesting dynamics. In practice, the mutation rate is set about 2
orders of magnitude less than the critical amount needed for
adaptation. In terms of speciation rates, the migration rate time
constant might correspond to something of the order of $10^4$ years,
instead of the 10 years or so one gets from considerations of doubling
times.

This code is released as EcoLab 3.5. Due to a design flaw, performance
of this code scales poorly with diversity, unless the code is run in
parallel with one cell per execution thread. For this experiment, the
runs took place on a $2\times2$ spatial grid, on a four processor
parallel computer supplied by the Australian Centre for Advanced
Computing and Communications, apart from one run of a $3\times3$ grid
on a 9 processor system. Work is currently underway to implement a
spatial version of the \EcoLab{} 4.x code, which does not suffer from
this performance problem.

\begin{figure}
  \epsfxsize=\figwidth\epsfbox{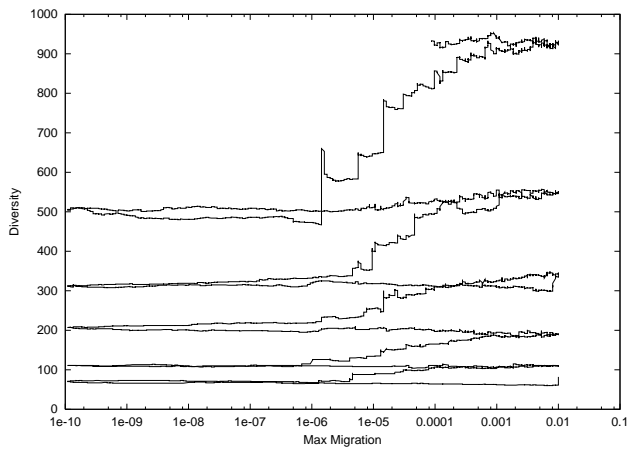}
\caption{Typical run sweeping the maximum migration rate $|\bgamma|_\infty$. The migration operator
  was applied every 100 time steps, so the units of the $x$-axis are
  0.01 cells per timestep}
\label{mig-sweep}
\end{figure}

The results of a typical run is shown in figure \ref{mig-sweep}. The
run started with a maximum migration rate of 0.01 at the bottom right
hand corner of the figure and swept down to $10^{-10}$ before
increasing. The migration rate was swept back and forwards 5 times
over the 18 million time steps in the run.

The first thing to note was that the expected response of diversity
to migration rate was not there. We would expect a response of the
form $D\propto A^{c}$, with $A=4$ in the $2\times2$ case, and $c$
varying smoothly between 1 for the infinite migration (panmictic) case
and 2 for zero migration. These results tentatively indicate that
possibly $c$ does not vary smoothly at all, but is nearly constant for
most values of $|\bgamma|_\infty$. This needs to be resolved with
further study.

\begin{figure}
  \epsfxsize=\figwidth\epsfbox{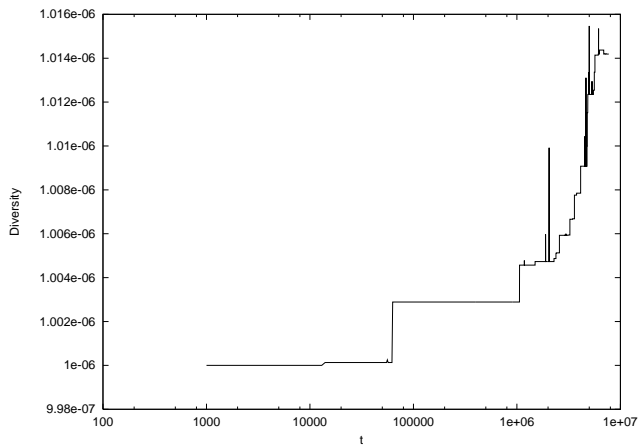}
\caption{Diversity growth on a $2\times2$ grid, with $|\bgamma|_\infty=1\times10^{-6}$.  }
\label{resonance}
\end{figure}

The second thing to note is the completely unexpected ``resonance'' at
about $1\times10^{-5}$. It is not peculiar statistical aberration,
since the same result was obtained with completely different random
number seeds, and fixing the migration rate at the resonance value
produces an exponential growth in diversity (Figure \ref{resonance}).

3 more tests were performed to determine if this result is an artifact
of discretisation, or a feature of the dynamics. The first involved
changing the grid to a $3\times3$ grid, which did not affect the
location of the resonance. The second involved scaling all parameters
in the model (\br, \bbeta, \bmu) by 0.1, which is equivalent to
changing the timescale. If the effect was purely due to dynamics, one
would expect the resonance to shift one order of magnitude higher on
the scale, however little qualitative different was observed. The
third test involved performing the migration operator every 1000
timesteps, instead of 100. This did change the resonance value by 1
order of magnitude, ruling out certain classes of software faults.

\section{Conclusion}

The choice of diversity as a proxy measure for ecosystem complexity is
a good choice. Complexity is obviously constrained by diversity, so
that bounded diversity dynamics also implies bounded complexity
dynamics. However, in the case of evolutionary Lotka-Volterra
dynamics, the system will tend to
self-organise\index{self-organization} to a critical surface where
speciation is balanced by extinction. This surface defines the maximum
allowed complexity for a given diversity value, which turns out to be
proportional to the diversity. The analysis presented in this paper
could be extended to other evolutionary ecologies as well.

Whilst there is still debate about whether the biosphere is exhibiting
unbounded complexity growth, I am persuaded by
Benton's\citeyear(2001){Benton01} argument that the growth is nothing
short of spectacular. In this paper I examined two possible mechanisms
for diversity growth --- {\em specialisation} which proves capable of
delivering unbounded creative evolution in \EcoLab, and {\em
  biogeographic provincialism}. Whilst I was only expecting
biogeographic changes to deliver a modest impact on diversity,
\EcoLab{} delivered a unexpected result of a ``resonance'', where if
the migration rate was tuned to this value, diversity grew
exponentially.

\section*{Acknowledgements}

I would like to thank the {\em Australian Centre for Adavanced
  Computing and Communications} for the computing resources needed to
  carry out this project.


\end{document}